# A Digital Twin-Based Simulation Framework for Safe Curve Speed Estimation Using Unity

**Araf Rahman**[*]
Ph.D. Student
Glenn Department of Civil Engineering
Clemson University, Clemson, South Carolina, 29634
Email: arafr@clemson.edu

**M Sabbir Salek, Ph.D.**
Senior Engineer
Glenn Department of Civil Engineering
Clemson University, Clemson, South Carolina, 29634
Email: msalek@clemson.edu

**Mashrur Chowdhury, Ph.D., P.E.**
Eugene Douglas Mays Chair of Transportation
Glenn Department of Civil Engineering
Clemson University, Clemson, South Carolina, 29634
Email: mac@clemson.edu

**Wayne A. Sarasua, Ph.D.**
Professor
Glenn Department of Civil Engineering
Clemson University, Clemson, South Carolina, 29634
Email: sarasua@clemson.edu

Word Count: 6,999 words + 2 table (250 words per table) = 7,499 words

**ABSTRACT**
Horizontal curves are often associated with roadway crashes due to speed misjudgment and loss of control. With the growing adoption of autonomous and connected vehicles, the accurate estimation of safe speed at curves is becoming increasingly important. The widely used AASHTO design method for safe curve speed estimation relies on an analytical equation based on a simplified point mass model, which often uses conservative parameters to account for vehicular and environmental variations. This paper presents a digital twin-based framework for estimating safe speed at curves using a physics-driven virtual environment developed in the Unity engine. In this framework, a real-world horizontal road curve is selected, and vehicle speed data are collected using a radar gun under various weather conditions. A 3D model of the road curve is constructed in a Unity environment using roadway geometric and elevation data. A parameterized vehicle model is integrated, allowing for variations in mass, acceleration, and center of gravity to reflect different vehicle types and loading scenarios. This simulation identifies the maximum safe speed at which a vehicle can traverse the given curve, providing a more vehicle and environment-specific estimate of the safe operating speed. The study validated that the safe curve speed estimates generated by the simulation were consistent with the real-world speed values observed at a curve. This study demonstrates how a physics-based digital twin can estimate a safer and more adaptive operating speed for vehicles traversing horizontal curves.

**INTRODUCTION**
       Horizontal curves on roads and highways are often associated with higher crash frequency and severity compared to tangent sections (*1*). In the United States, horizontal curves account for around 25% of the 38,000 fatal crashes that annually occur on highways (*2*). These segments experience an average crash rate that is three times higher than that of straight roadway segments (*3*). Similarly, roadway departure crashes are 1.5 to 4 times more likely to occur at horizontal curves compared to tangent sections (*4*). Identifying the hazards of driving on horizontal curves and mitigating them is a big part of achieving the goal of Vision Zero (*5*). These statistics underscore the critical need to identify and mitigate the specific risks associated with driving on curved sections, particularly in the context of Vision Zero goals for zero roadway fatalities (*5*).
       Numerous studies have identified several key factors contributing to crashes on horizontal curves, particularly those related to driver perception and environmental conditions. For example, the likelihood of crashes increases when a driver's undivided attention to the primary driving task is compromised by distractions caused by secondary in-vehicle tasks or, more critically, due to external environmental factors such as reduced visibility and wet road surfaces (*6*). The accurate perception of curvature or sharpness of the approaching curved segments necessitates that drivers modify their speed and maintain proper lane position to minimize the likelihood of accidents (*7*). The driver's failure to recognize or accurately perceive the sharpness of the upcoming curve may lead to run-off-road type crashes (*8*). These studies demonstrate the critical need for a driver to accurately perceive the sharpness of an approaching curve and estimate the safe curve negotiation speed.
       While an accurate perception of curve sharpness is critical, safe speed selection ultimately depends on the vehicle entering and traversing a curve at a speed that accounts for both roadway geometry and real-time conditions. Traditional methods for determining the safe speed on a curve, such as the American Association of State Highway and Transportation Officials (AASHTO) lateral acceleration model (*9*), use equations that use a simplified point mass system to represent the vehicle and assume fixed, conservative values for road friction. While these models are convenient and safe, their determined speeds are often very conservative. As a result, many drivers will drive at speeds higher than the design speed on curves, which in turn becomes a challenge to the enforcement of posted curve regulations (*1, 10*). In response to these limitations, researchers have proposed improved models that incorporate empirical data such as side friction demand,(*11, 12*) or dynamic signage based on observed speeds and crash data (*13*). However, these approaches miss out on important variables such as mass, suspension, and center of gravity. Curve speed warning systems (CSWS) have been developed to provide users with instantaneous feedback, by utilizing GPS, geometric and meteorological data to provide an alert to drivers when their speed has exceeded the safety threshold of an upcoming curve (*14, 15*); but these tools are still limited by the fixed parameters and oversimplified vehicle dynamics. These limitations have prompted a growing attention toward data-driven simulation frameworks that can accurately capture the complex interactions between vehicles and the road surfaces in a continuous adaptive manner.
      Digital Twins (DTs), envisioned initially by Dr. Michael Grieves in 2003 as a virtual replica of a physical system across its lifecycle, have seen growing use in autonomous vehicle (AV) development and road safety analysis. By enabling seamless data exchange between a physical entity and its digital counterpart, DTs support continuous simulation, predictive testing, and adaptive control in complex environments. For example, DTs have been employed to evaluate energy systems in AVs (*16*), incorporate real vehicles into virtual testing environments (*17*), and model harsh environmental conditions for AV behavior (*18*). These applications highlight the potential of DTs to simulate vehicle-road interactions in dynamic and customizable contexts, offering a promising alternative to simplified analytical models for estimating safe curve speed.
      Building on these foundational applications, DTs offer a powerful opportunity to rethink how safe speeds are estimated on curves. Traditional vehicle safety models, including those used in curve speed warning systems, typically rely on analytical equations with simplified assumptions about vehicle dynamics, friction, and road geometry. While computationally efficient, these models fall short when faced with real-time variability in vehicle behavior, loading conditions, or changing road surfaces. In

contrast, recent advances in DT simulation have demonstrated superior adaptability and predictive power. For instance, Liu et al. (*19*) integrated real-world traffic data with Long Short Term Memory(LSTM)-based learning and physics-driven simulation to dynamically assess highway safety more accurately than traditional risk models. Similarly, Dygalo et al. (*20*) implemented a DT to simulate braking systems in real time, overcoming the static limitations of conventional safety design. Further, "twin-in-the-loop" systems developed by Riva et al. (*21*) and Delcaro et al. (*22*) replace analytical estimators with continuously updating DTs, enabling better estimation of vehicle states under dynamically changing conditions. These developments suggest that DTs, especially those driven by real-time simulation, can serve as a robust platform for adaptive and context-specific curve speed estimation.

In this study, we present a DT-based simulation framework for estimating safe vehicle speed on horizontal curves using a high-fidelity, physics-driven model developed in the Unity engine. The model implements a six-degree-of-freedom (6-DOF) vehicle representation, featuring a sprung mass supported by four independently acting wheels, with steering inputs applied to the front axle. A real-world horizontal curve was selected, and vehicle speed data were collected on-site using a radar gun under various weather conditions. The simulation environment was built to replicate the geometry and elevation of the actual curve. By systematically varying vehicle parameters such as mass, center of gravity, and acceleration profiles, we identified the maximum safe speeds for different vehicle configurations. These simulated results were then validated against the real-world speed data to assess the accuracy of the model.

Our primary contributions are twofold: (1) the development of a customizable, physics-based DT for curve speed estimation, and (2) the empirical validation of this model using field data. This framework demonstrates the potential for real-time, vehicle and environment-specific curve speed advisories, contributing to more adaptive and safety-aware road systems.

## LITERATURE REVIEW

### Curve Safe Speed Estimation

Determining safe speeds for horizontal curves has remained a core challenge in transportation engineering and geometric design. Standard design practices, such as those proposed by AASHTO, define design speed based on roadway geometric features such as curvature, superelevation, and sight distance to maintain consistent safety regardless of vehicle class. However, studies such as that by Ervin et al. (*13*) have demonstrated that these methods overlook the complex interaction between roadway geometry and vehicle dynamics, particularly for heavy truck configurations, leading to stability issues even when operating with design limits. As a countermeasure, some agencies have introduced variable speed advisory signs that adjust posted speeds based on environmental cues. Yet, these solutions are still primarily based on fixed parameters and fail to account for differences in vehicle type.

Alternative estimation methods have worked towards more accurate predictions by incorporating empirical variables. For example, Pratt and Bonneson (*11*) introduced a curve speed estimation framework calibrated against real-world vehicle speeds collected across different visibility and congestion levels, using the driver's side friction demand as a key parameter. Similarly, Edquist et al. (*23*) carried out an extensive review of factors influencing driver speed selection, including roadway geometry, signage and environmental visibility. However, despite their improvements, these approaches fall short to capture individual vehicle dynamic parameters, such as the center of gravity, roll stiffness and lateral load transfer, which play a crucial role in curve safety.

Contemporary research has examined the shift from roadside signage systems to onboard warning technologies. The ISO 11067:2015 standard established the functional guidelines of Curve Speed Warning Systems (CSWS), which issue warnings to drivers when approaching curves at unsafe speeds. More advanced applications of CSWS utilize GPS data and curvature prediction algorithms to determine safe speeds (*24*, *25*). Lusetti et al. (*26*) introduced one of the early comprehensive models that accounts for the combined effects of driver behavior, vehicle characteristics and roadway features, representing a major shift. This led to advancements that integrated human factors into speed modelling, with systems that adjust with driver behavior, age, or preference (*27*).

Nonetheless, even advanced approaches like these often depend on parameter tuning or empirical rules and do not offer the real-time responsiveness required for precise estimation of safe curve speed. Studies by Bosetti et al. (*28*), Lee et al. (*29*), and Sun et al. (*30*) highlight sustained efforts to tailor speed advisory systems using vehicle-specific features like center of gravity or wheel span, or behavioral data, yet emphasize ongoing difficulties in calibration accuracy. Ultimately, many of these approaches highlight the growing shift toward simulation-based and data-driven methods that can accurately reproduce the complex, real-time interaction of vehicle, driver and environment on curved road segments.

**DTs in Connected and Autonomous Vehicle Systems**
The deployment of DT technology in the field of connected and autonomous vehicles (CAVs) has seen rapid growth in recent years, though its definitions and implementations remain diverse. As noted by Schwarz and Wang (*31*), there is significant conceptual overlap between DTs, model-based design and cyber-physical systems, especially within transportation. While many initial DT implementations centered on virtual modelling or Internet-of-Things-based data collection, newer approaches integrate cloud communication, edge-level processing, and artificial intelligence for live vehicle tracking and automated control.

Alam and El Saddik (*32*) introduced a theoretical DT architecture for driver assistance platforms supported by cloud infrastructure, emphasizing features, such as sensor fusion, computation, and control. Expanding upon this work, Kumar et al. (*33*) and Chen et al. (*34*) developed DT models aimed at predicting driver behavior, traffic flow, and inter-vehicular interactions leveraging 5G connectivity and machine learning algorithms. These systems seek to improve safety through inter-vehicle exchange of driver profiles, thereby enhancing the ability to respond to nearby traffic behavior.

The idea of parallel driving introduced by Wang et al. (*35*, *36*) transitions into cyber-physical-social systems (CPSS), where real-time vehicular data was transferred to a simulated environment for multi-agent interaction modelling. Building on these concepts, real-world DT implementations have also emerged. Wang et al. (*37*) developed a DT-supported Advanced Driver Assistance System (ADAS) that captured real-time vehicle data through vehicle-to-cloud communication, enabling instantaneous feedback loops and adaptive vehicle control. This DT-based approach was field-tested by Liao et al. (*38*), where it exhibited enhanced ramp merging behavior using the shared DT inputs. Similarly, Wang et al. (*39*) embedded DTs in a Unity-based simulation environment, creating a scalable simulation framework for connected vehicle applications.

These studies indicate that while DTs have mainly focused on data exchange and synchronization in CAV systems, their capabilities can be expanded to more high-resolution simulations of vehicle-infrastructure interactions.

**Physics-Based DT Simulation for Vehicle Dynamics**
Beyond coordinating and simulating driver behavior, DTs are being increasingly applied to simulate the way vehicles physically respond to variable driving conditions, which is a key requirement in safety-critical applications like curve speed estimation. These physics-based DTs simulate physical forces and vehicle-environment interactions, offering predictive insights that are not possible with simplified mathematical models.

Dygalo et al. (*20*) employed a combined digital and physical modeling environment to simulate brake behavior at the individual wheel level, facilitating real-time safety adaptation. Similarly, Liu et al. (*19*) implemented a DT framework that integrates LSTM-based learning with physics-based simulation to predict collision likelihood on highways, using live traffic video data. These models enabled trajectory prediction and dynamic safety control of diverse road conditions.

The "twin-in-the-loop" concept, explored by Riva et al. (*21*) and Delcaro et al. (*22*), presents a novel method in which traditional estimators, such as Kalman filters, are substituted with complete DT-based simulations. These simulators continuously determine and refine vehicle state estimates and demonstrate superior performance to traditional techniques in terms of accuracy and responsiveness. In a comparable use case, Wang et al. (*40*) implemented a human-in-the-loop DT simulation in the Unity engine, which tunes adaptive cruise control to match driver behavior. This demonstrates Unity's capability to adapt to the diverse needs of complex physical modelling.

The aforementioned studies highlight the growing role of DTs in real-time physics simulations and safety assessments. This shift enables a new generation of DT applications, that estimate safe curve speeds using physics-based simulations of vehicle-road dynamics under diverse conditions. These developments demonstrate the unexplored potential of DTs to substitute traditional fixed parameter models with adaptable vehicle-specific simulation approaches. This study contributes to this paradigm by developing and validating a DT framework in Unity to determine curve safe speed under dynamic vehicle and environmental factors.

**METHODS**

This study presents a DT-based simulation framework for curve safe speed estimation. A DT consists of two parts –physical assets, existing in physical space and their virtual twins, existing in digital space. The physical space sends data to the digital space to keep it up-to-date with the physical asset's latest state, while the digital space sends predictive analysis data to the physical space to improve its operation. **Figure 1** illustrates the DT framework presented in this study. Here, the physical assets include both the real-world horizontal curve being studied and the assortment of vehicles that will take the curve. On the virtual side, these physical assets translate to a 3-dimensional recreation of the horizontal curve and representative vehicle models corresponding to real vehicles. The digital space receives data regarding the vehicle type and its current speed. It then runs the simulation in real-time, adjusting it continuously using the live speed data. The digital space also simultaneously sends safe speed data from the simulation to the vehicle, completing the data loop. This study primarily focuses on creating the DT-based simulation. This can be summarized in three key stages: (1) selection of the study curve and real-world vehicle speed and class observation, (2) development of the DT of horizontal curve, and (3) development of the DT of vehicle models in Unity. The physics simulation is simply an integration of the virtual curve and vehicle models. The following sections elaborate on each of the stages.

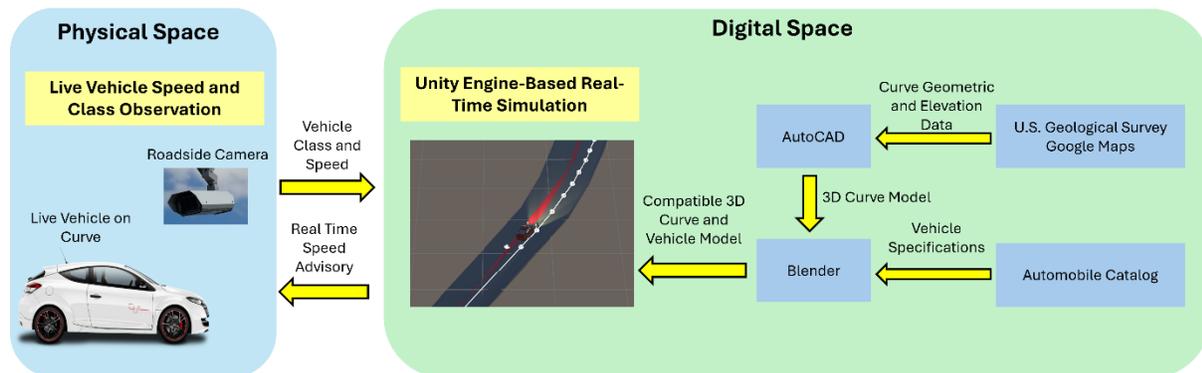

**Figure 1 DT-Based Horizontal Curve Simulation Framework**

**Study Curve Selection and Vehicle Speed Data Collection**
For this study, the study area chosen was Clemson, South Carolina. Horizontal curves around Clemson were surveyed based on traffic flow and curve radius. Based on the values obtained, curves were short-listed into two categories: high and low traffic flow. A reconnaissance survey of the high traffic flow curves was conducted in Google Maps Street View to assess lane configuration, potential data collection points and the nature of the surrounding terrain. A curve with a two-lane configuration was chosen, as it discards lane-changing, which has its own effect on speeds, and thus, would otherwise interfere with the effects of vehicle-road interactions, which are the focus of the study. The curve is located at Old Stadium

Road in Clemson, SC, featuring a radius of 712 feet, a superelevation of 7.8%, and a length of around 1,400 feet. The statutory posted speed limit at this curve is 35 mph. **Figure 2** shows the top-down view of the horizontal curve in Google Maps.

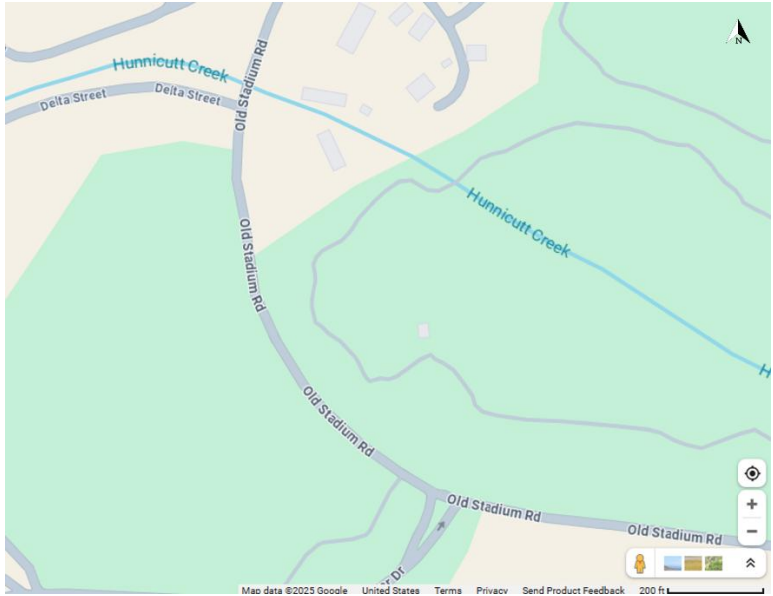

**Figure 2 Top-down view of horizontal curve on Old Stadium Road**

      Vehicle speed data were collected using a Bushnell radar gun, positioned slightly back from the curve's midpoint. The slightly off-midpoint positioning allowed the gun to accurately record the speeds of vehicles in and around the midpoint. Three speed observations were recorded for each vehicle – one at the midpoint and one each, slightly before and slightly after the midpoint. The mean was used as the mid-speed of the vehicle. In addition to the speed, the vehicle class was also recorded. The vehicles were classed into the following categories: (1) sedans, (2) sports utility vehicles (SUVs), (3) minivans, (4) pickup trucks, (5) single-unit large trucks, and (6) buses. Data collection was conducted during: (1) dry weather conditions, (ambient temperature: 67-71°F) and (2) wet weather conditions (rainfall intensity: 0.5 in/h). For each weather condition, a minimum sample size of 400 vehicles was observed. Data collection was performed during weekdays between 12:00 and 16:00 to capture typical daytime driving behavior. It was found that sedans, SUVs and pickup trucks accounted for around 90% of all vehicles observed, each accounting for roughly one-third of the traffic flow. Because of the large amount of recorded speed data, these three vehicle classes were selected for the simulation. From the 400 data points, the highest recorded curve speed for each vehicle class was defined as the maximum safe curve speed for that vehicle class.

**3D Reconstruction of a Horizontal Curve**
The selected curve was digitally constructed in the Unity engine using a multi-stage process beginning with comprehensive data collection. Geometric data about the curve's horizontal alignment was collected from satellite images on Google Maps. Elevation data at 1-meter resolution was extracted from USGS Digital Elevation Models (DEM), and this was used to determine the superelevation (7.8%) of the curve. This superelevation value was checked against field measurements and found to be accurate. Other roadway attributes, such as lane width (11 ft) and shoulder width (4 ft), were captured from Google Maps. The geometric data was used to construct a 3D model in AutoCAD. The tangent runout length and superelevation runoff length were determined from the AASHTO Green Book design standards (*9*), as the 1-meter resolution of the USGS DEM was insufficient to reliably detect these finer geometric transitions.

*Rahman et al.*

Similarly, 80 percent of the superelevation development was placed before the point of curvature in accordance with the AASHTO design standards (*9*). The AutoCAD model was imported into Blender and then exported from Blender in Filmbox (FBX) file format, which is compatible with Unity. This FBX file was imported into Unity. A route, consisting of a series of waypoints, was then placed along the curve to serve as guidance for the driving system.

**Unity Vehicle Model**
A parametrized vehicle simulation model was developed in Unity, leveraging its built-in physics engine together with user-defined components to capture realistic vehicle behavior. The vehicle body was modeled as a RigidBody object, enabling the specification of physical parameters, such as mass, aerodynamic drag, and center of mass. Each wheel was created using Unity's built-in WheelCollider component, which simulates wheel-ground interactions with adjustable parameters for friction, suspension and damping. The tire-road friction behavior within the WheelCollider system is based on a simplified version of the Pacejka 'Magic Formula' tire model (*41*). This model dynamically adjusts the friction force of the tire based on the slip. When the tire slip increases, the friction force initially also increases until the maximum friction force is reached at a certain slip value. In Unity, the maximum friction force developed is called extremumValue, and the slip at which it is developed is termed extremumSlip. After this, further slip reduces the friction force, until after a certain slip value, the friction force stabilizes to a constant value. Again, in Unity, the final constant friction force developed is called asymptoteValue, and the slip at which it is developed is termed asymptoteSlip. The final parameter that Unity uses, is tire stiffness, which is a multiplier on top of friction force. **Table 1** details all the simulation parameters as used in Unity, including the forward and side friction values in dry and wet weather conditions. For dry conditions, the default parameter settings in Unity were used as they simulate realistic vehicle behavior. For the wet condition, the parameters were adjusted to account for a 40-45% reduction in friction, which has been shown in previous studies (*42*). Conceptually, this Unity configuration models a vehicle as a sprung mass supported by four wheels. The front wheels steer, while the drive power is assigned to either front, rear, or all wheels based on the drivetrain setup. This framework can be described by a six-degrees-of-freedom (6-DOF) vehicle dynamic model. This framework includes translational velocities in the body along the fixed $x$ (longitudinal velocity, $u$), $y$ (lateral velocity, $v$), and $z$ (vertical velocity, $w$) directions, as well as rotational motion about all three axes: roll ($\varphi$), pitch ($\theta$), and yaw ($\psi$), respectively. The corresponding angular velocities around these axes are denoted as $p$, $q$, and $r$, respectively. **Figure 2** illustrates all six translational and rotational motions possible in the 6-DOF vehicle model. The equations presented in **Equations 1** through **6** express the 6-DOF vehicle dynamic model (*43*):

$$m(\dot{u} + qw - rv) = F_x + mg \cdot \sin\theta \quad (1)$$

$$m(\dot{v} + ru - pw) = F_y - mg \cdot \sin\alpha \cdot \cos\theta \quad (2)$$

$$m(\dot{w} + pv - qu) = F_z - mg \cdot \cos\alpha \cdot \cos\theta \quad (3)$$

$$I_x \dot{p} - (I_y - I_z)qr = M_x \quad (4)$$

$$I_y \dot{q} - (I_z - I_x)pr = M_y \quad (5)$$

$$I_z \dot{r} - (I_x - I_y)pq = M_z \quad (6)$$

Here, $m$ denotes the vehicle mass, while $I_x$, $I_y$, and $I_z$ represent its moments of inertia about the x-, y-, and z-axes respectively. The resultant forces along the vehicle's principal axes are $F_x$, $F_y$, and $F_z$, and the resultant moments about the corresponding axes are $M_x$, $M_y$, and $M_z$. **Equations 1** through **3**, which govern the vehicle's linear motion along the $x$, $y$, and $z$ axes, also account for gravitational force

*Rahman et al.*

components induced by road gradient. The equations of angular motion (**Equations 4** through **6**) characterize the change of the vehicle's angular velocities under the influence of tire-generated torques and inertial coupling effects. Collectively, these six equations constitute a comprehensive simulation framework for modelling vehicle behavior considering different variations of speed, mass distribution, and road banking.

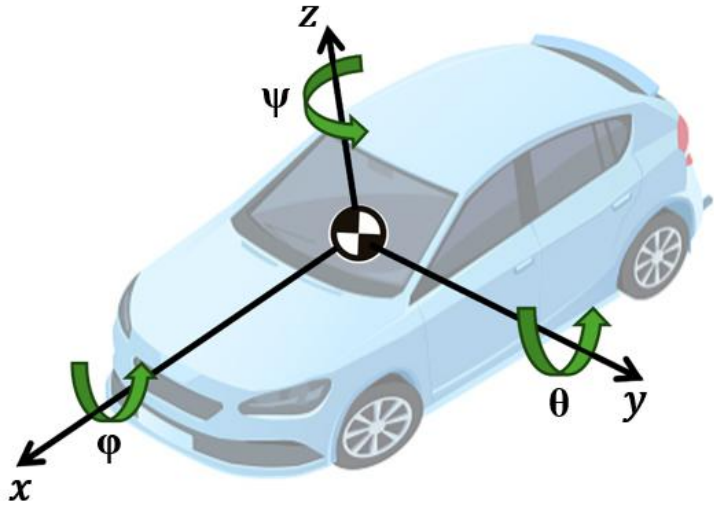

**Figure 2 All possible translational and rotational motions in a 6-DOF vehicle model**

**TABLE 1 Unity Simulation Parameters**

| Simulation Parameters | | | | |
|---|---|---|---|---|
| **Friction Parameters** | **Longitudinal Tire Characteristics** | | **Lateral Tire Characteristics** | |
| | Dry Pavement | Wet Pavement | Dry Pavement | Wet Pavement |
| extremumSlip | 0.40 | 0.40 | 0.40 | 0.40 |
| extremumValue | 1.00 | 0.65 | 1.00 | 0.65 |
| asymptoteSlip | 0.80 | 0.80 | 0.80 | 0.80 |
| asymptoteValue | 0.50 | 0.35 | 0.50 | 0.35 |
| stiffness | 1.00 | 0.85 | 1.00 | 0.85 |
| | | | | |
| **Vehicle Parameters** | **Sedan** | | **SUV** | **Pickup Truck** |
| Year, make and model | 2016 Chevrolet Malibu L 1.5 T | | 2015 Chevrolet Suburban E85 | 2015 Ford F-150 3.5-liter EcoBoost V6 |
| Vehicle mass (lbs) | 3286 | | 5864 | 5090 |
| Wheelbase (in) | 111.4 | | 130 | 145 |
| Overall length (in) | 193.8 | | 224.4 | 231.9 |
| Overall width (in) | 73.0 | | 80.5 | 79.9 |
| Overall height (in) | 57.7 | | 74.4 | 77.2 |

*Rahman et al.*

| Vehicle Parameters | Sedan | SUV | Pickup Truck |
|---|---|---|---|
| Center of gravity height (in) | 22 | 32 | 33 |
| Drive type | FWD | RWD | 4WD |
| Number of axles | 2 | 2 | 2 |

Three vehicle types were modelled in Unity based on the vehicle classes selected during data collection: 1) sedan, 2) SUV, and 3) pickup trucks. One specific real-world vehicle was chosen to represent each vehicle class. Prefabricated vehicle models corresponding to each vehicle type were downloaded from Unity's Asset Store and first modified in Blender to reflect the dimensions and wheelbase of real-world vehicles. They were then imported to Unity, where all subsequent modifications were made. The specifications of different vehicle models are readily available on Automobile Catalog, a vehicular database (*44*). Since the center of gravity location of different vehicle models was not readily available, an average value was chosen from a range for each class and assigned to the vehicle models (*45*). Similarly, for suspension, the default values set in Unity were used, as they simulate normal vehicle behavior (*41*). **Table 1** illustrates all the different real-world vehicles chosen for each class and their respective parameters as configured in Unity.

A car-controller script incorporating separate components to model the engine, transmission and steering, was developed to control vehicle movements. The engine of each vehicle type was modelled in Unity using the torque-speed curve of the respective engine models (*44*). Similarly, for each vehicle model, the transmission was modelled as a series of gears connected to the engine (*44*). A simplified speed-based gear shifting logic was implemented to model the vehicle's acceleration at different speeds. The torque was delivered to WheelColliders, collected as a list. The torque delivery and WheelCollider list were modified to account for different drivetrains such as Rear-Wheel Drive (RWD), Front-Wheel Drive (FWD) or All-Wheel Drive (AWD) for the respective vehicles. An autonomous steering system was implemented to follow the waypoint-based route placed along the curve. The route was superimposed on the outer lane of the curve, as the maximum real vehicle speed recorded was for the outer lane. The route was formulated such that it would follow a path that deviated 1.25 ft from the lane centerline towards the inner edge of the lane. This was done because a previous study has shown that drivers tend to cut the curve by slightly over 1 ft on average, on the outer lanes of rural roads (*46*). The entire route was placed in the car-controller script as a list of waypoints, reflecting their order along the route. **Figure 3** shows a portion of the curve in Unity with the route placed. The script iterates over the list and selects a pair of target waypoints based on the order in the list and a configurable threshold distance along the direction of the route between the front of the vehicle and the waypoints. Once the target waypoint pair is identified, the front wheels clip to the angle between the waypoint pair, with a damping rate, in the simulation. The front wheels maintain this angle until the distance between the waypoint of the target pair nearest to the vehicle's front decreases below the threshold distance. After this, the script identifies the next target waypoint pair, and the process continues until the last waypoint in the route is reached. The car controller of the script also incorporates a brake. To model actual vehicle behavior along the curve, the complete route was partitioned into five distinct sections, as follows: normal driving, curve entry, full superelevation, curve exit, and post-curve acceleration. This partition reflects the field-observed speed variation patterns along horizontal curves, as shown by Figueroa Medina and Tarko (*47*), who found that drivers often decelerate and accelerate non-uniformly through the curve. Each segment was also configured with characteristic throttle and brake inputs to emulate acceleration and deceleration phases observed in the aforementioned study (*47*), with throttle multipliers scaling the final torque output based on the driving phase, enabling more realistic simulation of vehicles traversing curves. **Table 2** illustrates the 5 discrete curve segments and their respective throttle multipliers and brake responses used in this study. Two speeds on the curve, $v_{base}$ and $v_{curve}$, can be configured, which correspond to the tangent speed and curve speed, respectively. For each vehicle, $v_{curve}$ was incrementally increased until the respective vehicle was no longer able to negotiate the curve safely. This was taken as the maximum safe speed for that vehicle and condition. The $v_{base}$ was adjusted to be 5 mph in excess of the $v_{curve}$ speeds.

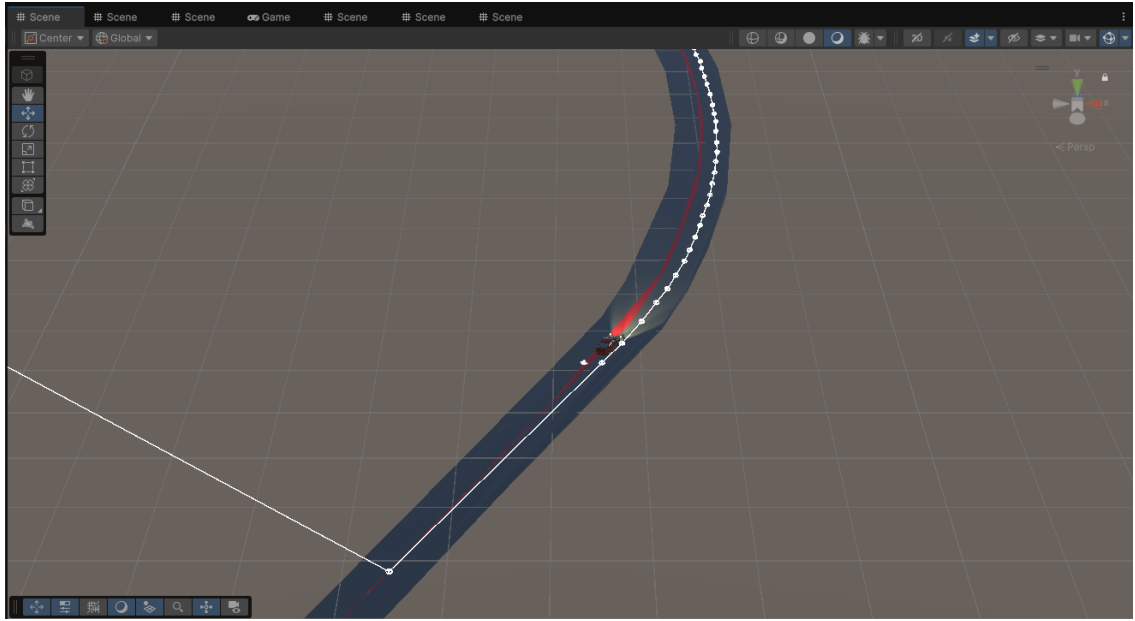

**Figure 3 Horizontal curve in Unity with route placed on top**

**TABLE 2 Vehicle Control Behavior Across Curve Phases**

| Driving Phase | Target Speed | Throttle Multiplier | Braking Behavior | Description |
|---|---|---|---|---|
| Normal Driving | Base speed ($v_{base}$) | 1.0 if $v <$ $v_{base}$, else 0.0 | Brake if $v > v_{base} +$ 2 mph | Vehicle accelerates to base cruising speed with full throttle |
| Curve Entry | Curve speed ($v_{curve}$) | 0.0 | Braking only after driver reaction delay ($\tau_r = 1.21$s) | Throttle cut; initiates delayed braking for safe curve negotiation |
| Full Superelevation | Curve speed ($v_{curve}$) | 0.1 if $v <$ $v_{curve}$, else 0.0 | Brake if $v > v_{curve} +$ 1 mph | Gentle throttle to hold curve speed; minor braking for excess speed |
| Curve Exit | Base speed ($v_{base}$) | Constant 0.1 | None | Moderate throttle applied to begin acceleration out of curve |
| Post-Curve Acceleration | Base speed ($v_{base}$) | 1.0 if $v <$ $v_{base}$, else 0.0 | None | Full acceleration resumes once curve is exited |

## RESULTS AND DISCUSSIONS

The DT-based simulation model built in Unity, was used to determine the maximum safe curve speed for three vehicle categories, i.e., sedan, SUV and pickup truck, under both dry and wet pavement conditions. For all vehicle types, the Unity model estimated safe turning speeds that facilitated safe curve maneuvering while maintaining traction and rollover resistance. The Unity-defined safe speeds were compared with the real-world observed maximum speeds. The AASHTO design method was used as a benchmark method to compare the performance of the Unity simulation. When the AASHTO design method was used to determine maximum safe speed, the AASHTO design speed was considered the maximum safe speed.

**Figures 4** and **6** illustrate the maximum recorded safe speeds for dry and wet pavement conditions, respectively, for each vehicle type in Unity, the corresponding maximum recorded real-world speed data, and the AASHTO-defined safe speed for the corresponding class. It can be observed from **Figure 4** that under dry conditions, the Unity-defined safe curve speed is closer to the observed real maximum speeds than the AASHTO-defined safe speed for all three vehicle classes considered in this study. This is consistent with the fact that the AASHTO-defined safe speed uses conservative friction parameters, and in the real world, there is often a higher friction supply than the AASHTO demand, allowing drivers to go at speeds higher than the intended design speed. This fact is also confirmed in **Figure 5**, which depicts the percentage deviations of the Unity-defined safe speed and AASHTO-defined design speed from the observed real-world speeds. Here, the Unity-defined maximum safe speed consistently has a lower percentage deviation compared to the AASHTO-defined one. In **Figure 6**, a slightly different picture can be observed for the case of wet pavements. Here, the AASHTO-defined safe speeds are closer to the observed speeds for two cases (sedans and pickup trucks), while the Unity-defined safe speed is closer to the observed speed for SUVs. It must be noted that in both cases, i.e., for sedan and pickup trucks, the Unity-defined safe speeds were lower than those observed. Past research has shown that, on the outer lanes of rural roads, drivers will travel closer to the inner edge, allowing for higher speeds (*46*). In this study, the route chosen was such that it was shifted 1.25 ft from the lane centerline to the inner edge. This shift was chosen based on the average value observed for drivers on the outer lane of rural roads (*46*). However, the same study also showed that drivers may travel up to 3 ft from the lane centerline to the inner edge, for an 11 ft lane (*46*). Thus, the maximum observed speeds were likely from the drivers who were taking a tighter turn at a higher speed. **Figure 7** presents the percentage deviations of the AASHTO-defined safe speed and Unity-defined maximum safe speed from the observed speed. The Unity-defined maximum safe speed performed worse than the AASHTO-defined safe speed only for the case of pickup trucks.

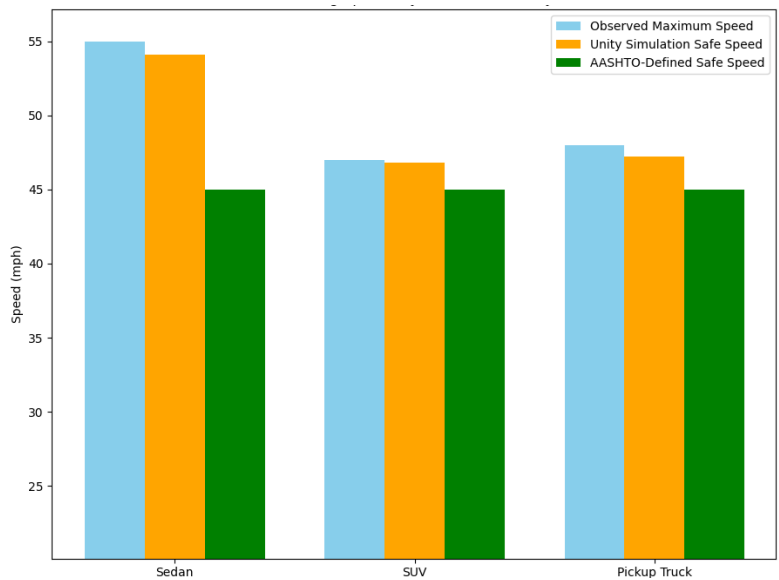

**Figure 4 Safe curve speed for each vehicle type for dry conditions**

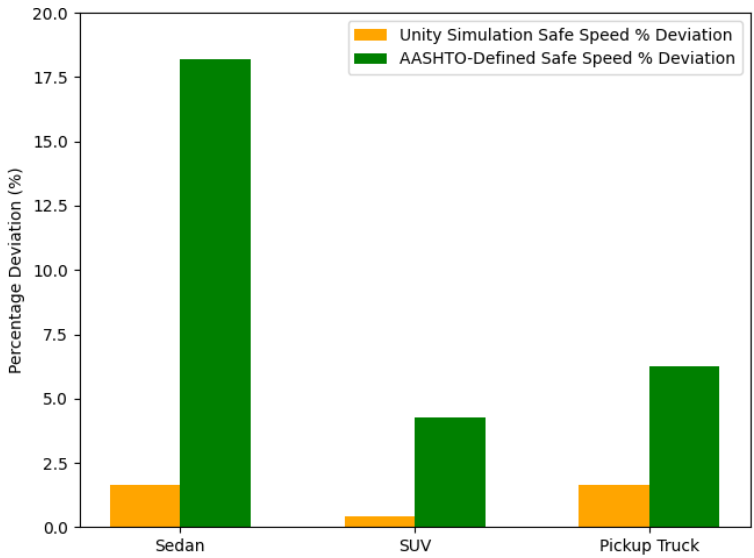

**Figure 5  Percentage deviation of Unity and AASHTO-defined safe speed from real observed maximum speed for dry conditions**

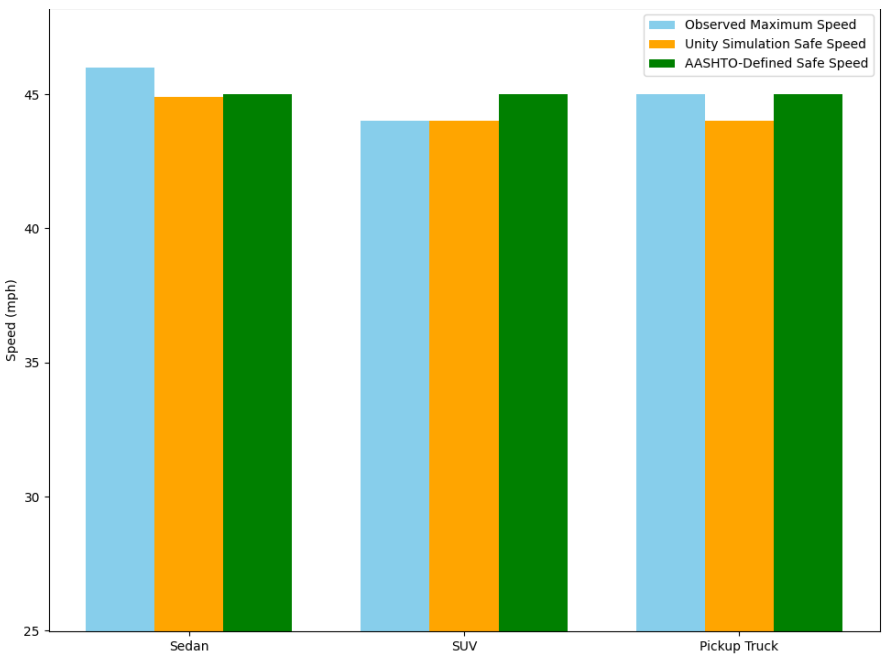

**Figure 6 Safe curve speed for each vehicle type for wet conditions**

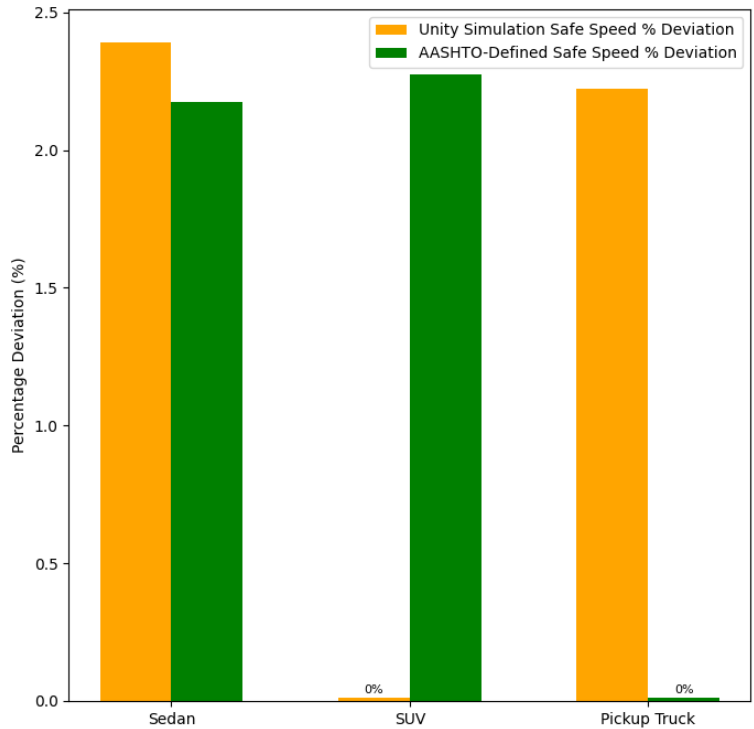

**Figure 7 Percentage deviations of Unity and AASHTO-defined safe speed from real observed maximum speed for wet conditions**

**CONCLUSIONS**

    This study presented a DT-based framework for predicting safe curve speeds for vehicles navigating horizontal curves using Unity's physics engine for vehicle modelling. The DT framework was used to simulate vehicles traversing a curve, and it provided safe speeds specific to each vehicle and road surface condition. The maximum safe speeds provided by the DT framework and the AASHTO design speed of the curve were both compared with the highest real-world speeds observed on the curve, to see how much each deviated from the highest observed speed. In dry road conditions, the DT framework provided speeds were closer to the observed speeds, than the AASHTO design speed. In wet conditions, the DT only showed considerably higher deviation than the AASHTO design for the pickup truck, but for the other two vehicle types it had similar or lower deviation. The results demonstrate the capability of the framework to accurately estimate safe speeds of different vehicles on a pre-determined horizontal curve under different pavement conditions. One of the limitations of the DT framework is that it is sensitive to driver behavior and route placement, which are difficult to quantify. While this study focuses on a single curve and three vehicle classes, the framework can be expanded to other vehicle classes and curves. Similarly, the framework could also be extended to multi-vehicle and multi-lane scenarios using car-following models and lane-changing models. In practical deployment scenarios, the DT could operate either on edge devices or on the onboard units of connected and autonomous vehicles. The follow-up study will explore the scalability of the system and its computational efficiency, which are critical aspects of real-world deployment.

**ACKNOWLEDGMENTS**

    This work is based upon the work supported by the Center for Regional and Rural Connected Communities ($CR^2C^2$) (a U.S. Department of Transportation Region 4 University Transportation Center) headquartered at North Carolina A&T State University, Greensboro, North Carolina, USA. Any opinions, findings, conclusions, and recommendations expressed in this material are those of the author(s) and do

not necessarily reflect the views of $CR^2C^2$, and the U.S. Government assumes no liability for the contents or use thereof

**AUTHOR CONTRIBUTIONS**
The authors confirm contribution to the paper as follows: study conception and design: A. Rahman, M.S. Salek, M. Chowdhury; data collection: A. Rahman; analysis and interpretation of results: A. Rahman; draft manuscript preparation: A. Rahman, M.S. Salek, M. Chowdhury, W. A. Sarasua. All authors reviewed the results and approved the final version of the manuscript.
    Moreover, the authors acknowledge using large language models (LLMs), particularly OpenAI's ChatGPT, to make editorial improvements to this manuscript. LLM was solely used for refining grammar and formatting texts and references. No information, figure, table or analysis has been generated using LLMs.

**DECLARATION OF CONFLICTING INTERESTS**
The authors declared no potential conflicts of interest with respect to the research, authorship, and/or publication of this article.

# REFERENCES

1. Dhahir, B., and Y. Hassan. Using Horizontal Curve Speed Reduction Extracted from the Naturalistic Driving Study to Predict Curve Collision Frequency. *Accident Analysis & Prevention*, Vol. 123, 2019, pp. 190–199. https://doi.org/10.1016/j.aap.2018.11.020.

2. Othman, S., R. Thomson, and G. Lannér. Safety Analysis of Horizontal Curves Using Real Traffic Data. *Journal of Transportation Engineering*, Vol. 140, No. 4, 2014. https://doi.org/10.1061/(asce)te.1943-5436.0000626.

3. Wang, B., S. Hallmark, P. Savolainen, and J. Dong. Crashes and Near-Crashes on Horizontal Curves along Rural Two-Lane Highways: Analysis of Naturalistic Driving Data. *Journal of Safety Research*, Vol. 63, 2017, pp. 163–169. https://doi.org/10.1016/j.jsr.2017.10.001.

4. Donnell, E., J. Wood, S. Himes, and D. Torbic. Use of Side Friction in Horizontal Curve Design: A Margin of Safety Assessment. *Transportation Research Record*, Vol. 2588, No. 1, 2016, pp. 61–70. https://doi.org/10.3141/2588-07.

5. Achieving the Vision of Zero Roadway Deaths Through the Safe System Approach | US Department of Transportation. https://www.transportation.gov/utc/achieving-vision-zero-roadway-deaths-through-safe-system-approach. Accessed July 28, 2025.

6. Charlton, S. G. The Role of Attention in Horizontal Curves: A Comparison of Advance Warning, Delineation, and Road Marking Treatments. *Accident Analysis & Prevention*, Vol. 39, No. 5, 2007, pp. 873–885. https://doi.org/10.1016/j.aap.2006.12.007.

7. Reymond, G., A. Kemeny, J. Droulez, and A. Berthoz. Role of Lateral Acceleration in Curve Driving: Driver Model and Experiments on a Real Vehicle and a Driving Simulator. *Human Factors*, Vol. 43, No. 3, 2001, pp. 483–495. https://doi.org/10.1518/001872001775898188.

8. Shinar, D. Curve Perception and Accidents on Curves: An Illusive Curve Phenomenon? *Z VERKEHRSSICHERHEIT*, Vol. 23, No. 1, 1977, pp. 16–21.

9. Policy on Geometric Design of Highways and Streets (7th Edition), Including 2019 Errata - 3.3 Horizontal Alignment - Knovel. https://app.knovel.com/web/view/pdf/show.v/rcid:kpPGDHSE12/cid:kt0121N7H2/viewerType:pdf//root_slug:33-horizontal-alignment/url_slug:horizontal-alignment?cid=kt0121N7H2&b-toc-cid=kpPGDHSE12&b-toc-title=Policy%20on%20Geometric%20Design%20of%20Highways%20and%20Streets%20%287th%20Edition%29%2C%20including%202019%20Errata&b-toc-url-slug=horizontal-alignment&hierarchy=kt0121N6V2. Accessed July 28, 2025.

10. Wang, B., S. Hallmark, P. Savolainen, and J. Dong. Examining Vehicle Operating Speeds on Rural Two-Lane Curves Using Naturalistic Driving Data. *Accident Analysis & Prevention*, Vol. 118, 2018, pp. 236–243. https://doi.org/10.1016/j.aap.2018.03.017.

11. Pratt, M. P., and J. A. Bonneson. Assessing Curve Severity and Design Consistency Using Energy- and Friction-Based Measures. *Transportation Research Record*, Vol. 2075, No. 1, 2008, pp. 8–15. https://doi.org/10.3141/2075-02.

12. Bonneson, J. A., and M. P. Pratt. Model for Predicting Speed along Horizontal Curves on Two-Lane Highways. *Transportation Research Record*, Vol. 2092, No. 1, 2009, pp. 19–27. https://doi.org/10.3141/2092-03.

13. Ervin, R. D., C. C. MacAdam, and M. Barnes. INFLUENCE OF THE GEOMETRIC DESIGN OF HIGHWAY RAMPS ON THE STABILITY AND CONTROL OF HEAVY-DUTY TRUCKS. *Transportation Research Record*, No. 1052, 1986.

14. Jiménez, F., Y. Liang, and F. Aparicio. Adapting ISA System Warnings to Enhance User Acceptance. *Accident Analysis & Prevention*, Vol. 48, 2012, pp. 37–48. https://doi.org/10.1016/j.aap.2010.05.017.

15. Chowdhury, S., M. Faizan, and M. Hayee. Advanced Curve Speed Warning System Using Standard GPS Technology and Road-Level Mapping Information. Presented at the 6th International Conference on Vehicle Technology and Intelligent Transport Systems, Prague, Czech Republic, 2020.

16. Rassõlkin, A., T. Vaimann, A. Kallaste, and V. Kuts. Digital Twin for Propulsion Drive of Autonomous Electric Vehicle. Presented at the 2019 IEEE 60th International Scientific Conference on Power and Electrical Engineering of Riga Technical University (RTUCON), 2019.

17. Szalai, M., B. Varga, T. Tettamanti, and V. Tihanyi. Mixed Reality Test Environment for Autonomous Cars Using Unity 3D and SUMO. Presented at the 2020 IEEE 18th World Symposium on Applied Machine Intelligence and Informatics (SAMI), 2020.

18. Ge, Y., Y. Wang, R. Yu, Q. Han, and Y. Chen. Demo:Research on Test Method of Autonomous Driving Based on Digital Twin. Presented at the 2019 IEEE Vehicular Networking Conference (VNC), 2019.

19. Liu, J., Z. Tang, C. Liu, Y. Zhao, and X. Pan. Highway Driving Safety Analysis and Management Based on Digital Twin. Presented at the 2023 IEEE 3rd International Conference on Digital Twins and Parallel Intelligence (DTPI), 2023.

20. Dygalo, V., A. Keller, and A. Shcherbin. Principles of Application of Virtual and Physical Simulation Technology in Production of Digital Twin of Active Vehicle Safety Systems. *Transportation Research Procedia*, Vol. 50, 2020, pp. 121–129. https://doi.org/10.1016/j.trpro.2020.10.015.

21. Riva, G., S. Formentin, M. Corno, and S. M. Savaresi. Twin-in-the-Loop State Estimation for Vehicle Dynamics Control: Theory and Experiments. http://arxiv.org/abs/2204.06259. Accessed July 18, 2025.

22. Delcaro, G., F. Dettù, S. Formentin, and S. M. Savaresi. Automatic Dimensionality Reduction of Twin-in-the-Loop Observers. http://arxiv.org/abs/2401.10945. Accessed July 18, 2025.

23. Edquist, J., C. M. Rudin-Brown, and M. G. Lenné. ROAD DESIGN FACTORS AND THEIR INTERACTIONS WITH SPEED AND SPEED LIMITS.

24. Chen, X. L., M. Lu, D. Z. Zhang, R. Chi, and J. Wang. Integrated Safety Speed Model for Curved Roads. 2010.

25. Li, K., H.-S. Tan, J. A. Misener, and J. K. Hedrick. Digital Map as a Virtual Sensor – Dynamic Road Curve Reconstruction for a Curve Speed Assistant. *Vehicle System Dynamics*, Vol. 46, No. 12, 2008, pp. 1141–1158. https://doi.org/10.1080/00423110701837110.

26. Lusetti, B., L. Nouveliere, S. Glaser, and S. Mammar. Experimental Strategy for a System Based Curve Warning System for a Safe Governed Speed of a Vehicle. Presented at the 2008 IEEE Intelligent Vehicles Symposium, 2008.

27. Glaser, S., S. Mammar, and C. Sentouh. Integrated Driver–Vehicle–Infrastructure Road Departure Warning Unit. *IEEE Transactions on Vehicular Technology*, Vol. 59, No. 6, 2010, pp. 2757–2771. https://doi.org/10.1109/TVT.2010.2049670.

28. Bosetti, P., M. Da Lio, and A. Saroldi. On Curve Negotiation: From Driver Support to Automation. *IEEE Transactions on Intelligent Transportation Systems*, Vol. 16, No. 4, 2015, pp. 2082–2093. https://doi.org/10.1109/TITS.2015.2395819.

29. Lee, Y. H. Deng, Rochester Hills, MI (US) (73) Assignee: GM Global Technology Operations,.

30. Sun, C., C. Wu, D. Chu, M. Zhong, Z. Hu, and J. Ma. Risk Prediction for Curve Speed Warning by Considering Human, Vehicle, and Road Factors. *Transportation Research Record*, Vol. 2581, No. 1, 2016, pp. 18–26. https://doi.org/10.3141/2581-03.

31. Schwarz, C., and Z. Wang. The Role of Digital Twins in Connected and Automated Vehicles. *IEEE Intelligent Transportation Systems Magazine*, Vol. 14, No. 6, 2022, pp. 41–51. https://doi.org/10.1109/MITS.2021.3129524.

32. Alam, K. M., and A. El Saddik. C2PS: A Digital Twin Architecture Reference Model for the Cloud-Based Cyber-Physical Systems. *IEEE Access*, Vol. 5, 2017, pp. 2050–2062. https://doi.org/10.1109/ACCESS.2017.2657006.

33. Kumar, S. A. P., R. Madhumathi, P. R. Chelliah, L. Tao, and S. Wang. A Novel Digital Twin-Centric Approach for Driver Intention Prediction and Traffic Congestion Avoidance. *Journal of Reliable Intelligent Environments*, Vol. 4, No. 4, 2018, pp. 199–209. https://doi.org/10.1007/s40860-018-0069-y.

34. Chen, X., E. Kang, S. Shiraishi, V. M. Preciado, and Z. Jiang. Digital Behavioral Twins for Safe Connected Cars. Presented at the MODELS '18: ACM/IEEE 21th International Conference on Model Driven Engineering Languages and Systems, Copenhagen Denmark, 2018.

35. Wang, F.-Y. Parallel Control and Management for Intelligent Transportation Systems: Concepts, Architectures, and Applications. *IEEE Transactions on Intelligent Transportation Systems*, Vol. 11, No. 3, 2010, pp. 630–638. https://doi.org/10.1109/TITS.2010.2060218.

36. Wang, F.-Y., N.-N. Zheng, D. Cao, C. M. Martinez, L. Li, and T. Liu. Parallel Driving in CPSS: A Unified Approach for Transport Automation and Vehicle Intelligence. *IEEE/CAA Journal of Automatica Sinica*, Vol. 4, No. 4, 2017, pp. 577–587. https://doi.org/10.1109/JAS.2017.7510598.

37. Wang, Z., X. Liao, X. Zhao, K. Han, P. Tiwari, M. J. Barth, and G. Wu. A Digital Twin Paradigm: Vehicle-to-Cloud Based Advanced Driver Assistance Systems. Presented at the 2020 IEEE 91st Vehicular Technology Conference (VTC2020-Spring), 2020.

38. Liao, X., Z. Wang, X. Zhao, K. Han, P. Tiwari, M. J. Barth, and G. Wu. Cooperative Ramp Merging Design and Field Implementation: A Digital Twin Approach Based on Vehicle-to-Cloud Communication. *IEEE Transactions on Intelligent Transportation Systems*, Vol. 23, No. 5, 2022, pp. 4490–4500. https://doi.org/10.1109/TITS.2020.3045123.

39. Wang, Z., K. Han, and P. Tiwari. Digital Twin Simulation of Connected and Automated Vehicles with the Unity Game Engine. Presented at the 2021 IEEE 1st International Conference on Digital Twins and Parallel Intelligence (DTPI), 2021.

40. Wang, Z., K. Han, and P. Tiwari. Digital Twin Simulation of Connected and Automated Vehicles with the Unity Game Engine. Presented at the 2021 IEEE 1st International Conference on Digital Twins and Parallel Intelligence (DTPI), 2021.

41. Technologies, U. Unity - Manual: Wheel Collider Component Reference. https://docs.Unity3d.com/6000.1/Documentation/Manual/class-WheelCollider.html. Accessed July 17, 2025.

42. Watanabe, A., I. Kageyama, Y. Kuriyagawa, T. Haraguchi, T. Kaneko, and M. Nishio. Study on the Influence of Environmental Conditions on Road Friction Characteristics. *Lubricants*, Vol. 11, No. 7, 2023, p. 277. https://doi.org/10.3390/lubricants11070277.

43. Vehicle Body 6DOF - Two-Axle Vehicle Body with Translational and Rotational Motion - Simulink. https://www.mathworks.com/help/vdynblks/ref/vehiclebody6dof.html. Accessed July 21, 2025.

44. Zal, P., MEng, and MBA. Automobile-Catalog the Catalog of Cars, Car Specs Database. https://www.automobile-catalog.com/#gsc.tab=0. Accessed July 17, 2025.

45. Gillespie, T. D. *Fundamentals of Vehicle Dynamics*. 1992.

46. Fitzsimmons, E. J., S. S. Nambisan, R. R. Souleyrette, and V. Kvam. Analyses of Vehicle Trajectories and Speed Profiles Along Horizontal Curves. *Journal of Transportation Safety & Security*, Vol. 5, No. 3, 2013, pp. 187–207. https://doi.org/10.1080/19439962.2012.680573.

47. Figueroa Medina, A. M., and A. P. Tarko. Speed Changes in the Vicinity of Horizontal Curves on Two-Lane Rural Roads. *Journal of Transportation Engineering*, Vol. 133, No. 4, 2007, pp. 215–222. https://doi.org/10.1061/(asce)0733-947x(2007)133:4(215).